\documentclass[twocolumn,showpacs,preprintnumbers,amsmath,xamssymb,aps]{revtex4-1}

\usepackage{amssymb,amsfonts,amstext,graphics,graphicx,subfigure}

\usepackage{mathtools}
\usepackage{setspace}
\usepackage[colorlinks]{hyperref}
\usepackage{color}

\usepackage{mathbbol}

\def\XXint#1#2#3{{\setbox0=\hbox{$#1{#2#3}{\int}$ }
\vcenter{\hbox{$#2#3$ }}\kern-.5\wd0}}

\def\calc{\mathcal{C}}

\def\R{\mathbb{R}}

\def\Z{\mathbb{Z}}

\def\bq{\begin{equation}}
\def\eq{\end{equation}}
\def\bqy{\begin{eqnarray}}
\def\eqy{\end{eqnarray}}

\def\tensI{\underline{\underline{I}}}

\def\tensGa{\underline{\underline{\Gamma}}}

\def\de{\delta}
\def\De{\Delta}
\def\ep{\epsilon}

\def\Ga{\Gamma}

\def\bfB{\mathbf{B}}

\def\bfE{\mathbf{E}}

\def\bfJ{\mathbf{J}}
\def\bfP{\mathbf{P}}
\def\bfM{\mathbf{M}}

\def\bfL{\mathbf{L}}
\def\bfR{\mathbf{R}}

\def\bfk{\mathbf{k}}
\def\bfx{\mathbf{x}}
\def\bfX{\mathbf{X}}
\def\bfv{\mathbf{v}}

\def\p{\partial}

\def\and{\quad\mathrm{and}\quad}

\def\dmu{\, d^3x  d^3v}

\newcommand{\abs}[1]{\ensuremath{\left \vert #1 \right \vert}}

\begin{document}

\title{Algebra of Invariants for the Vlasov-Maxwell System}
\author{Philip J.~Morrison}
\email{morrison@physics.utexas.edu}
\affiliation{Department of Physics and Institute for Fusion Studies, 
The University of Texas at Austin, Austin, TX, 78712, USA}
\date{\today}

\begin{abstract}

The algebra of invariants for both the relativistic and nonrelativistic multispecies Vlasov-Maxwell system is examined, including the case with a fixed ion background.  Invariants and  their associated fluxes are obtained directly from the Vlasov-Maxwell system.  The invariants are shown to Poisson commute with the Hamiltonian and the rest of the Poisson bracket algebra of invariants is identified.  Special attention is given to the role played by the monopole condition, 	$\nabla\cdot\bfB$.

\bigskip

{\it Dedicated to the memory of Robert L.\ Dewar,  treasured friend and colleague.}
 
\end{abstract}

\maketitle

 \tableofcontents

\section{Introduction}
\label{sec:intro}

In this paper we consider constants of motion for the multispecies Vlasov-Maxwell system of equations with particular emphasis on the role played by the constraints of Gauss's law  and the condition that the magnetic field be solenoidal.  The usual method for obtaining the energy-momentum tensor and angular momentum is by means of an action principle and Noether's theorem,  as was done, e.g.,  for the Vlasov-Maxwell system in \cite{pjmP85}.  However,  an alternative is to proceed with the gauge-free Hamiltonian structure of the Vlasov system in terms of the  noncanonical  Poisson bracket   \cite{pjm80,pjm82,MW,ibb2}, which is done here.   

Although early gauge-free field theoretic structures and generalized Poisson brackets appeared in the physics literature  \cite{pauli33,pauli53,martin59,bornin,sudarshan,ibb},  the first occurrence in the plasma literature  appeared in \cite{pjmG80}  and  in \cite{pjm82,pjm98,pjm05}, which  can be consulted for review.  

It was first pointed out in \cite{pjm82}  that the role of the Gauss's law and the solenoidal magnetic field constraints appear in the Hamiltonian formulation on different footing:  the solenoidal magnetic field is necessary for the Jacobi identity, an essential property  of the Poisson bracket of a Hamiltonian formulation, while Gauss's law appears as a true Casimir invariant, an invariant associate with degeneracy of a Poisson bracket.  Thus monopole theories are not Hamiltonian field theories.  (See \cite{pjm13,pjmH18} for further details.)  This distinction between the constraints is carefully tracked as we investigate constants of motion and their associated algebra of invariants.  Throughout an eye is kept out on the possible ramifications of the use of invariants in the context of numerical simulation.

The paper is organized as  follows.  In Sec.~\ref{sec:vlasov-maxwell} the Vlasov-Maxwell system, its noncanonical Hamiltonian structure, and the associated Jacobi identity and collection of Casimir invariants are reviewed.  Next, in Sec.~\ref{sec:DandF} the ten conservation laws for the full relativistic Vlasov-Maxwell system along with its Casimir invariants are directly obtained, including their associated fluxes.   The role played by the  constraints in leading to invariance is explicitly delineated.  Similarly the invariants of the nonrelativisic theory are explored.  In Sec.~\ref{sec:AoI} the algebra of invariants, the pairwise Poisson brackets of all invariants are calculated. It is shown that the Poisson bracket with invariants produces a realization of the algebra associated with the Poincar\'e group in the relativistic case and the Euclidean group in the nonrelativistic case.  Section \ref{sec:discussion} contains discussion of the consequences of preceding sections. In particular, in Sec.~\ref{ssec:frozen} it  is shown that the usual system that freezes ion motion does not conserve invariants, notably  momentum. Computational consequences are discussed in Sec.~\ref{ssec:consequences} and a system that describe the nonlinear dynamics of the weibel instability are discussed in Sec.~\ref{ssec:weibel}. Next,  in Sec.~\ref{ssec:assymetry} possible ramifications of the asymmetry of the constraints are mentioned.   The paper concludes in Sec.~\ref{sec:summary} with a summary.

\section{The Vlasov-Maxwell System}
\label{sec:vlasov-maxwell}

\subsection{Equations of Motion}
\label{ssec:eom}

The relativistic Vlasov equation for a particle species $s$,  charge $q_s$, and mass $m_s$ is 
\bqy
\label{Rvlasov}
&& \frac{\p f_s}{\p t} +  \frac{\bfv}{\sqrt{1+ \abs{\bfv}^{2}}}  \cdot \nabla f_s 
\\
&& \hspace{ 1.5 cm}+ \frac{q_s}{m_s} \bigg(\bfE  + \frac{\bfv}{\sqrt{1+ \abs{\bfv}^{2}}}  \times \bfB\bigg) \cdot \frac{\p f_s}{\p \bfv} = 0.
\nonumber
\eqy
while the nonrelativistic version is given by 
\bq
\label{NRvlasov}
\frac{\p f_s}{\p t} + \bfv \cdot \nabla f_s + \frac{q_s}{m_s} \left(\bfE + \bfv \times \bfB\right) \cdot \frac{\p f_s}{\p \bfv} = 0\,.
\eq
In \eqref{Rvlasov} and \eqref{NRvlasov} $f_s$ denotes the phase space density, which is a function of $\bfx,\bfv$ and $t$. 

Both the relativistic and nonrelativistic cases are  coupled nonlinearly to  Maxwell equations,
\bqy
&& \frac{\p\bfE}{\p t} - \nabla \times \bfB = - \bfJ\,, 
\label{ampere}
\\
&&\frac{\p \bfB}{\p t}  +  \nabla \times \bfE = 0\,, 
\label{faraday}
\\
&&
\nabla\cdot\bfE = \rho\,,  \quad\mathrm{and}\quad 
\nabla\cdot \bfB = 0\,,
 \label{constraints}
\eqy
with the  sources  in the relativistic case   given by  
\bqy
 \rho &=& \sum_s q_s \int f_s \, d^3v \quad \mathrm{and}
\nonumber\\  
\bfJ &=& \sum_s q_s
 \int \frac{\bfv}{\sqrt{1+ \abs{\bfv}^{2}}} \,  f_s\, d^3v\,, 
\label{relS}
\eqy
which together constitute a four-vector, while in the nonrelativistic case  they are given by 
\bqy
\rho &=& \sum_s q_s \int f_s \, d^3v  \quad \mathrm{and}
\\
\hat\bfJ &=& \sum_s q_s \int  \bfv \, f_s\, d^3v\,.
\label{NrelS}
\eqy

\subsection{Noncanonical Hamiltonian Structure}
\label{sec:NCbkt}

The  Hamiltonian form of equations of motion is
\bq
\frac{\p \Psi}{\p t} =\{\Psi, H\}\,,
\label{eom}
\eq
where for the Vlasov-Maxwell theories the dynamical variables are represented by $\Psi=(f_s,\bfE,\bfB)$.  Thus given a Poisson bracket $\{\,,\,\}$ and a Hamiltonian $H$ the equations of motion are defined. 

The Vlasov-Maxwell  noncanonical Poisson bracket, which is the same for both the relativistic and nonrelativistic theories,   is given by   
\bqy
\{F,G\}  
&=& \sum_{s}\frac1{m_s} \int \left[ \frac{\de F}{\de f_s}  \frac{\de G}{\de f_s} \right] {f_s} \, d^3x d^3v
\nonumber \\
&+& \sum_{s} \frac{q_s}{m_s^2} \int  \bfB \cdot \left( \frac{\p}{\p \bfv}  \frac{\de F}{\de f_s} \times 
\frac{\p}{\p \bfv} \frac{\de G}{\de f_s} \right) f_s \, d^3x d^3v 
\nonumber \\
 && \hspace{-1 cm} +\sum_{s} \frac{q_s}{m_s} \int  \left(
        \frac{\p}{\p \bfv} \frac{\de F}{\de f_s} \cdot \frac{\de G}{\de \bfE}
      -  \frac{\p}{\p \bfv}  \frac{\de G}{\de f_s} \cdot \frac{\de F}{\de \bfE}
   \right) f_s\,d^3x d^3v
   \nonumber \\
&+& \int \left( \frac{\de F}{\de \bfE} \cdot \nabla \times \frac{\de G}{\de \bfB} - \frac{\de G}{\de\bfE} \cdot \nabla \times \frac{\de F}{\de\bfB} \right) d^3x\, .
\label{VMbracket}
\eqy
where 
\bq
[f,g]\coloneqq  \nabla f\cdot  \frac{\p g}{\p \bfv} - \nabla g\cdot  \frac{\p f}{\p \bfv} \,.
\eq

With the relativistic Hamiltonian,  
\bqy
H  &=&  \sum_{s}   {m_s} \int\sqrt{1+ \abs{\bfv}^{2}} \,  f_s  \, d^3  x d^3  v 
\nonumber\\
&& \hspace{ 2cm}
+   \frac{1}{2} \int \big( \abs{\bfE}^{2} + \abs{\bfB}^{2} \big)\,  d^3x\,,
\label{Rham}
\eqy
the form of \eqref{eom}, with Poisson bracket \eqref{VMbracket}, yields \eqref{Rvlasov},  \eqref{ampere}, and \eqref{faraday} with the $\bfJ$ of \eqref{relS}, while with the nonrelativistic Hamiltonian,
\bqy
\hat{H}  &=& \sum_{s}   \frac{m_s}{2 }\int  \abs{\bfv}^{2}   f_s \, d^3  x d^3  v
\nonumber\\
&& \hspace{ 2cm}
 +   \frac{1}{2} \int \big( \abs{\bfE}^{2} + \abs{\bfB}^{2} \big) \, d^3x\,,
\label{NRham}
\eqy
\eqref{eom} yields \eqref{NRvlasov},  \eqref{ampere}, and \eqref{faraday} with the $\hat\bfJ$ of \eqref{NrelS}.

When there is a need to make distinction between relativistic and nonrelativistic quantities, we  will use the hat symbol $\, \hat{\ }\,$ to denote nonrelativistic quantities. 
 
\subsection{Jacobi identity and Casimir invariants}
\label{sec:casimirs}

In order for the formalism described in Sec.~\ref{sec:NCbkt} to actually be a Hamiltonian formalism, it is necessary for the bracket of \eqref{VMbracket} to statisfy the Jacobi identity (see, e.g.,  \cite{pjm82,pjm98}),  
\bq
 \{\{F,G\},H\} + \{\{G,H\},F\}  +  \{\{H,F\},G\} =0\,, 
 \eq
 for all functionals $F,G,H$. The early calculation of  \cite{pjm82} (see \cite{pjm13} for explicit details) revealed  
 \bqy
&& \{\{F,G\},H\} + \mathrm{cyc} =
\label{MVjac}\\
&&\sum_s\!\frac{q_s}{m_s^3 }\!\int     \!\nabla\cdot \bfB\,  \, 
 \frac{\p}{\p \bfv}  \frac{\de F}{\de f_s}\! \cdot
 \left(  \frac{\p}{\p \bfv}  \frac{\de G}{\de f_s}\! \times \!\frac{\p}{\p \bfv}  \frac{\de H}{\de f_s} \right)
 f_s\,d^3x \,d^3v\,.
\nonumber
\eqy
Thus the Jacobi identity is not satisfied   unless it is restricted to  functionals defined on divergence-free magnetic fields.  Observe, that of the two constraints of \eqref{constraints} the Jacobi identity only relies on $\nabla\cdot\bfB=0$ and not on  $\nabla\cdot\bfE=\rho$.   Thus the  gauge invariant Hamiltonian structure with the Poisson bracket of \eqref{VMbracket} only requires $\bfB$ interpreted as a 2-form to be closed, but it need not be exact. 

 Casimirs are constants of motion that  originate from the degeneracy of the Poisson bracket and, consequently,  they are invariants for any Hamiltonian.   They are functionals that satisfy  $\{ C, \mathcal{G} \} = 0$ for any functional $G$.

For the Vlasov-Maxwell bracket several such Casimirs  $C(f, \bfE, \bfB)$ are known \cite{pjm87,pjmP89,pjmCGBT13}.  First, for a arbitrary function $\calc_{s}$ of  $f_{s}$ the following is a Casimir:
\bq
C_{s} = \int \calc_{s} (f_{s} ) \, \dmu \,.
\label{liouC}
\eq
This family of Casimirs is a manifestation of Liouville's theorem and corresponds to conservation of phase space volume \cite{pjm87}.
Next, there are  two Casimirs related to constraints of \eqref{constraints},
\begin{align}
	C_{E} &= \int h_{E}(\bfx)\,  \big(  \nabla\cdot \bfE - \rho \big)\,  d^3 x\,  ,
	\label{CE} \\
C_{B} &= \int h_{B}(\bfx) \,  \nabla\cdot \bfB \,  d^3 x\,  ,
\label{CB}
\end{align}
where $h_{E}$ and $h_{B}$ are arbitrary  functions of $\bfx$.  The  functional $C_{B}$ is not a true Casimir because  it must be identically zero for the Jacobi identity to be satisfied.  For this reason it is sometimes called a  pseudo-Casimir. It acts like a Casimir in that its Poisson bracket with any other functional vanishes, but it must be born in mind that the Jacobi identity is only satisfied when $ \nabla\cdot \bfB = 0$.  In contrast, dynamics with an initial condition $ \nabla\cdot \bfE - \rho\neq 0$ is still Hamiltonian dynamics.   This distinction followed from the direct (and tedious) Jacobi identity calculation of Ref.~\onlinecite{pjm82}, written out explicitly in an appendix of \cite{pjm13}.

\section{Invariant densities and fluxes}
\label{sec:DandF}

In this section we calculate directly from the equations of motion the conservation laws for densities and their corresponding fluxes.  A conservation law has the form
\bq
\frac{\p R}{\p t} + \nabla\cdot  \boldsymbol{\Ga}=0
\eq
for a density $R$  with associated  flux $\boldsymbol{\Ga}$.  Similarly, if $\bfR$ is a vector density, such as momentum,  we have
\bq
\frac{\p \bfR}{\p t} + \nabla\cdot \tensGa=0
\eq
where $\tensGa$ is the corresponding tensor flux.  Observe, if the flux, vector or tensor,  through a boundary vanishes, then
we have the constants of motion, 
\bq
\frac{d}{dt} \int R\, d^3x=0 \quad\mathrm{or}\quad  \frac{d}{dt} \int \bfR\, d^3x=0 \,.
\label{COM}
\eq

For the case of a phase space density $\calc$ the conservation law is given by
\bq
\frac{\p \calc }{\p t} + \nabla\cdot \boldsymbol{\Ga}^x +\frac{\p }{\p  \bfv}\cdot \,  \boldsymbol{\Ga}^v =0
\eq
and similar to \eqref{COM} we have 
\bq
\frac{d}{dt} \int \calc\, \dmu=0\,.
\eq

 For the relativistic case we expect 10 constants of motion, corresponding to the 10 elements of the Poincar\'{e} group, which is appropriate for relativistically covariant theories. For the nonrelativistic case we lose invariants because it mixes the relativistic invariance of Maxwell's equations with a Gallilean invariant from of the Vlasov equation.  We will see that there are 7 invariants, 6 corresponding to invariants associated with the Euclidean group plus the Hamiltonian.

\subsection{Casimir Fluxes}

The density associated with the Casimir \eqref{liouC} has the conservation law for the relativistic Vlasov system
\bqy
&&
\frac{\p \calc_s}{\p t} +   \nabla\cdot\bigg(\calc_s  \frac{\bfv}{\sqrt{1+ \abs{\bfv}^{2}}} \bigg)
\\
&&\hspace{ 1.2 cm}
+ \frac{\p }{\p  \bfv}\cdot \, \left( \frac{q_s}{m_s}  \calc_s \, 
\bigg(\bfE +  \frac{\bfv}{\sqrt{1+ \abs{\bfv}^{2}}}  \times \bfB\bigg) \right) = 0\,, 
\nonumber
\eqy
while  for the nonrelativistic Vlasov system it has the form
\bq
\frac{\p \hat\calc_s}{\p t} +   \nabla\cdot(\hat\calc_s \bfv ) +
\frac{\p }{\p  \bfv}\cdot \, \left( \frac{q_s}{m_s} \hat\calc_s \, \big(\bfE + \bfv \times \bfB\big) \right) = 0\,. 
\eq

Consider $C_E$ of \eqref{CE},
\bq
\frac{\p \big(h_E(\nabla\cdot \bfE - \rho)\big)}{\p t} =h_E \left(-\nabla\cdot\bfJ - \frac{\p \rho}{\p t}\right) \equiv 0\,,
\eq
where for the relativistic case $\bfJ$ is given by \eqref{relS} and for the nonrelativistic  case by  $\hat\bfJ$ of \eqref{NrelS}.  This follows from   \eqref{ampere} and  \eqref{Rvlasov} or  \eqref{NRvlasov}. Thus there is no  flux associated with this Casimir. 
Similarly, because $C_B$ of \eqref{CB} only depends on Maxwell's equations, its conservation property is the same for both the relativistic and nonrelativistic cases.  From \eqref{faraday} it follows immediately that 
\bq
\frac{\p (h_B\nabla\cdot\bfB)}{\p t} =h_B\nabla\cdot\frac{\p \bfB}{\p t} \equiv 0\,.
\eq
Likewise,  there is no  flux associated with this Casimir.

\subsection{Conservation of momentum}

The relativistic and nonrelativistic Vlasov-Maxwell systems conserve  the same  total momentum 
\bq
\bfP=\sum_s m_s \int   \bfv\,  f_s \,    \dmu + \int   \bfE\times\bfB \, d^3x\,,
\eq 
which has the associated momentum vector density 
\bq
 \bfR^P= \sum_s  m_s \int   \bfv\,  f_s    \,   d^3v +   \bfE\times\bfB\,,
 \label{RP}
\eq
i.e.,  $\bfP=\int   \bfR^P\,   d^3x$.    To obtain the momentum flux tensor  $\tensGa^P$,  we calculate $\p \bfR^P/\p t$ and   insert \eqref{Rvlasov}, \eqref{faraday}, and \eqref{ampere}.  Then, upon making use of the vector identities
\bqy
  \bfB\times (\nabla\times\bfB)  &=&\nabla  {|\bfB|^2}/{2} - \bfB\cdot\nabla\bfB
\nonumber\\
 \bfB\cdot\nabla\bfB&=& \nabla\cdot (\bfB\otimes\bfB) -\bfB\ (\nabla\cdot\bfB)\,, 
 \nonumber
 \eqy
 we obtain
 \bq
    \frac{\partial  \bfR^P}{\partial t} + \nabla\cdot \tensGa^P =-\bfB (\nabla\cdot  \bfB) + \bfE\,  (\rho -\nabla\cdot  \bfE)\,, 
    \label{conP}
\eq
with 
\bqy
 \tensGa^P &=& \sum_s  m_s\int  \frac{\bfv \otimes\bfv}{\sqrt{1+ \abs{\bfv}^{2}}}  \,   f_s    \,   d^3v  
  \label{RPflux}\\
 && \hspace{ 1 cm}
 -   \bfE\otimes\bfE -  \bfB\otimes\bfB
+  \tensI \, \left( |\bfE|^2 +  |\bfB|^2\right)\!/2\,,
\nonumber
 \eqy
where $\tensI$ is the unit tensor.
 
 From \eqref{conP} it follows that 
 \bq
 \frac{d\bfP}{dt}=\int \big(  \bfE\,  (\rho -\nabla\cdot  \bfE) -\bfB\,  \nabla\cdot  \bfB\Big)d^3x=0\,, 
    \label{dP0}
\eq
where the first equality of \eqref{dP0}  uses a  no flux boundary condition (e.g.\   periodicity) and the second follows if    Gauss's law   is satisfied and $\nabla\cdot\bfB=0$.

 Similarly, for the nonrelativistic theory we obtain the momentum flux 
 \bqy
 \hat{\tensGa}^P &=& \sum_s m_s  \int  {\bfv \otimes\bfv}  \,   f_s    \,   d^3v
 \label{Pflux} \\
 &&\hspace{1.5cm}
  -   \bfE\otimes\bfE -  \bfB\otimes\bfB
+  \tensI \, \left( |\bfE|^2 +  |\bfB|^2\right)/2\,.
\nonumber
 \eqy
 
\subsection{Conservation of energy}

From \eqref{Rham} it is evident that the relativistic energy density is
\bq
R^H  = \sum_{s}   {m_s} \int\sqrt{1+ \abs{\bfv}^{2}} \,  f_s  \,   d^3 v +   \frac{1}{2}   \big( \abs{\bfE}^{2} + \abs{\bfB }^{2} \big)\,,
\label{RH} 
\eq
 while from \eqref{NRham}  the nonrelativistic energy density is seen to be,
\bq
\hat{R}^H  = \sum_{s}   \frac{m_s}{2} \int  \abs{\bfv}^{2} \,  f_s   \,   d^3 v +   \frac{1}{2}   \big( \abs{\bfE}^{2} + \abs{\bfB }^{2} \big)\,. 
\eq
As for the momentum flux, we proceed directly to obtain 
\bq
    \frac{\partial  R^H}{\partial t} + \nabla\cdot  \boldsymbol{\Ga}^H =0\,, 
    \label{conH}
\eq
where we identify 
 \bq
 \boldsymbol{\Ga}^H= \bfR^P=  \sum_s  m_s  \int  \bfv\,  f_s    \,   d^3v +   \bfE\times\bfB\,,
 \eq
using \eqref{RP}. 

For the nonrelativistic theory the calculation yields
 \bq
 \hat{\boldsymbol{\Ga}}^H =  \sum_s \frac{m_s}{2}  \int  |\bfv|^2 \, \bfv\,  f_s    \,   d^3v +   \bfE\times\bfB\,.
 \eq
 
\subsection{Conservation of center of mass}
\label{ssec:Ccom}

Consider the quantity
\bq
\bfM=  \int  R^H \bfx \, d^3x - \bfP \, t\,  ,
\label{bfM}
\eq
with the associated vector density
\bqy
\bfR^M &=&  R^H \bfx -\bfR^P t
\nonumber\\
&=&  \bigg(\sum_{s} \!  {m_s}\! \int\! \sqrt{1+ \abs{\bfv}^{2}} \,  f_s  \,   d^3 v 
+   \frac{1}{2}   \big( \abs{\bfE}^{2} + \abs{\bfB }^{2} \big)\!\bigg)\, \bfx
\nonumber\\ 
&-& \bigg(\sum_s  m_s \int \!  \bfv\,  f_s    \,   d^3v +   \bfE\times\bfB\bigg)\,  t
\label{RM}\,.
\eqy
Clearly, up to boundary terms, 
\bqy
\frac{d \bfM}{dt}&=&  \int  \frac{\p R^H}{\p t} \bfx \, d^3x -\bfP - \frac{d \bfP}{dt} t 
\nonumber\\
&=&-\int \bfx\, \nabla\cdot \bfR^P \, d^3x - \bfP = \bfP-\bfP=0\,, 
\eqy
assuming $d \bfP/dt=0$. 

Again, to obtain the flux we consider 
\bqy
\frac{\p \bfR^M}{\p t} &=& \frac{\p R^H}{\p t}\, \bfx- \bfR^P -\frac{\p \bfR^P}{\p t}\,  t  
\nonumber\\
&=&-  (\nabla\cdot \boldsymbol{\Ga}^H )\, \bfx - \bfR^P 
\nonumber\\
&+&\big(\nabla\cdot \tensGa^P +\bfB (\nabla\cdot  \bfB) - \bfE\,  (\rho -\nabla\cdot  \bfE)\big)\, t
\nonumber\\
&=& -\nabla\cdot\big(  \boldsymbol{\Ga}^H\! \otimes \bfx - \tensGa^P\,  t \big)
 \label{COMF}\\
&&\hspace{1.5 cm}+\, \big(\bfB (\nabla\cdot  \bfB) - \bfE\,  (\rho -\nabla\cdot  \bfE)\big)\, t\,.
\nonumber
\eqy
Thus, 
\bq
    \frac{\partial  \bfR^M}{\partial t} +  \nabla\cdot   \tensGa^M = 
     \big(\bfB (\nabla\cdot  \bfB) - \bfE\,  (\rho -\nabla\cdot  \bfE)\big)\, t\,, 
\label{COMM}
\eq
with
\bq
  \tensGa^M= \boldsymbol{\Ga}^H\! \otimes \bfx -  \tensGa^P\,  t  \,.
\eq

  In the nonrelativistic case, there does not appear to be an invariant of the type of $\bfM$.  this is likely a manifestation that the nonrelativistic Vlasov equation has neither full  Galilean nor Poincar\'{e} invariance,  while the Maxwell equations have Poincar\'{e} invariance.   Thus the  nonrelativistic theory  does not have either full group. 
 
\subsection{Conservation of angular momentum}

The total angular momentum is given by 
\bq
\bfL=\sum_s m_s \int \bfx\times  \bfv\,  f_s \,    \dmu + \int \bfx\times(\bfE\times\bfB)\, d^3x\,,
\eq 
with the density
\bq
\bfR^L=\sum_s m_s \int \bfx\times  \bfv\,  f_s \, d^3v +  \bfx\times(\bfE\times\bfB) \,,
\eq 
Direct calculation gives
\bqy
\frac{\p \bfR^L}{\p t}&=&
-\sum_s m_s \int (\bfx\times  \bfv)\,  \frac{\bfv}{\sqrt{1+ \abs{\bfv}^{2}}}\cdot\nabla f_s \,    d^3v
\nonumber\\
&& + \rho\,  \bfx\times\bfE + \bfx\times (\bfJ\times\bfB)
\nonumber\\
&&+   \bfx \times(\nabla |\bfE|^2/2 + \bfE\cdot\nabla \bfE)  
\nonumber\\
&& + \bfx \times(\nabla |\bfB|^2/2 + \bfB\cdot\nabla \bfB) 
- \bfx\times (\bfJ\times\bfB)\,.
\nonumber
\eqy
In order to simply this, the following are useful:
\bqy
\bfx\times\nabla h&=&-\nabla\times(\bfx h)
\nonumber\\
(\bfx \times (\bfB\cdot\nabla \bfB))_i&=&\ep_{ijk} x_j\big( \p_{\ell} (B_{\ell} B_k) - B_k \nabla\cdot \bfB \big)
\nonumber\\
&=&\ep_{ijk} \p_{\ell} (x_j  B_{\ell} B_k)  - \ep_{ijk} x_j B_k \nabla\cdot \bfB 
\nonumber\\
\ep_{ijk} x_j \frac{v_k v_{\ell}}{\sqrt{1+ \abs{\bfv}^{2} }}  \p_{\ell} f_s
&=&  \p_{\ell} \bigg( \ep_{ijk}  \frac{ x_jv_k v_{\ell}}{\sqrt{1+ \abs{\bfv}^{2} }}\,  f_s\bigg)\,,
\nonumber
\eqy
where $\p_\ell=\p /\p x_\ell$.
We obtain
 \bq
    \frac{\partial  \bfR^L}{\partial t} + \nabla\cdot \tensGa^L =-(\bfx\times \bfB) (\nabla\cdot  \bfB) +(\bfx\times  \bfE)\,  (\rho -\nabla\cdot  \bfE)\,, 
    \label{conL}
\eq
with the angular momentum flux tensor $\tensGa^L$ given by  
\bqy
\Ga^L_{\ell i}&=& \sum_s m_s \int  \ep_{ijk}  \frac{ x_jv_k v_{\ell}}{\sqrt{1+ \abs{\bfv}^{2} }}\,  f_s\, d^3v
\label{gammaL}\\
&-&\ep_{i j \ell} x_j  \, \left(E_\ell E_k +  B_\ell B_k\right)
+\ep_{i j \ell} x_j  \, \left( |\bfE|^2 +  |\bfB|^2\right)/2 \,.
\nonumber
\eqy

The nonrelativistic case has the flux
\bqy
\hat{\Ga}^L_{\ell i}&=& \sum_s m_s \int  \ep_{ijk}   x_jv_k v_{\ell} \,  f_s\, d^3v
\label{NgammaL}\\
&-& \ep_{i j \ell} x_j  \, \left(E_\ell E_k +  B_\ell B_k\right)
+\ep_{i j \ell} x_j  \, \left( |\bfE|^2 +  |\bfB|^2\right)/2 \,.
\nonumber
\eqy

 
 
\section{Algebra of Invariants}
\label{sec:AoI}

Poisson brackets, like that of \eqref{VMbracket}, together with the set of functionals on which it acts constitute a Lie algebra, in fact a Lie algebra realization on functionals.  A closed subset of functionals is a Lie subalgebra.  One such closed subset is the algebra of invariants, these are functionals that Poisson commute with the Hamiltonian, but need not  commute with each other, i.e.,  the algebra is not in general abelian.   Poisson's theorem, which follows from the Jacobi identity,  states that the Poisson bracket of any two invariants is also an invariant.  Therefore a closed algebra can be  obtained by calculating  the Poisson brackets of all invariants with each other.   
In this section we first show that the invariants obtained in Sec.~\ref{sec:DandF}  commute with the Hamiltonian $H$, then calculate their mutual Poisson brackets. 

\subsection{Relativistic Invariants}
\label{ssec:RI}

For the relativistic Vlasov-Maxwell theory we first calculate $\{ H  , \Phi \}$
with $H$ given by \eqref{Rham} and  $\Phi\in\{H,\bfP,\bfL, \bfM\}$.    Obviously  by antisymmetry, $\{ H  , H \}=0$. 
Before calculating the other brackets, for convenience we rewrite the invariants as follows:
\bqy
\bfP&=& \int \bfR^P\, d^3x 
\label{Pinv}\,,
\\
\bfL&=& \int \bfR^L\, d^3x=  \int \bfx\times\bfR^P\, d^3x\,,
\label{Linv}\\
\bfM&=& \int (R^H \bfx - \bfR^P t  )\, d^3x\,,
\label{Minv}
 \eqy
with $\bfR^P$ given by \eqref{RP} and  $R^H$  by \eqref{RH}.  When calculating  Poisson brackets it is convenient to write our invariants as, e.g.,  $\bfP=\bfP^K + \bfP^F$, where $\bfP^K$ is the kinetic part of $\bfP$ that depends on $f_s$ and   $\bfP^F$ is the part of $\bfP$ that depends  the fields  $\bfE$ and $\bfB$.  This is possible for all the invariants  because they are all the sum of  a  part that depends only on the fields $f_s$ part with a part that depends only on the fields $\bfE$ and $\bfB$.

For our calculations we assume boundary conditions that allow dropping  all boundary terms obtained from integration by parts. For example, decay conditions in velocity and periodic boundary conditions in space, achieve this goal. 
When calculating $\{ H  , \Phi \}$ the following will be needed:
\bq
 \frac{\de H}{\de f_s}=   m_s \sqrt{1+ \abs{\bfv}^{2}}\,,
\quad \frac{\de H}{\de \bfE}= \bfE
\,,\quad \frac{\de H}{\de \bfB}= \bfB\,.
\eq
Below, we  consider each invariant in  turn.

\medskip
\noindent\underline{$\{\bfP,H\}$}: 
\medskip

Using
\bq
\frac{\de P_i}{\de f_s}= m_sv_i\,,
\quad \frac{\de P_i}{\de E_\ell}= \ep_{i\ell t} B_t
\,,\quad \frac{\de P_i}{\de B_\ell}= \ep_{is\ell} E_s\,.
\label{Pde}
\eq
and  $\bfP^F= \int   \bfE\times\bfB \, d^3x$, 
we obtain from \eqref{VMbracket}, 
\bq
\{\bfP^K, H\}= \int (\rho \bfE  +\bfJ\times\bfB)\, d^3x
\label{PKH}
\eq
while 
\bq
\{\bfP^F, H\}=- \int (\bfE\,\nabla\cdot \bfE  + \bfB\,\nabla\cdot \bfB + \bfJ\times\bfB)\, d^3x\,.
\label{PFH}
\eq
Thus,  adding \eqref{PKH} and \eqref{PFH} gives
\bq
\{\bfP, H\}= \int \big( \bfE \,(\rho-\nabla\cdot\bfE)   -\bfB\,\nabla\cdot \bfB\big)\, d^3x\,,
\label{bktPH}
\eq
as expected from \eqref{conP}.

\medskip
\noindent\underline{$\{\bfL,H\}$}: 
\medskip

Using
\bqy
 \frac{\de L_i}{\de f_s}&=& m_s\ep_{ijk} x_jv_k\,,
 \\
\quad \frac{\de L_i}{\de E_\ell}&=&  \ep_{rik} \ep_{u\ell k} x_rB_u
= (\bfx\cdot\bfB) \de_{\ell i} - B_i x_{\ell}  \,,
\\
\quad \frac{\de L_i}{\de B_\ell}&=&  \ep_{rik} \ep_{\ell u k} x_rE_u = x_\ell E_i - (\bfx\cdot\bfE) \de_{\ell i} \,.
\eqy
we obtain 
\bq
\{\bfL, H\}= \int\Big(\bfx\times\big(\bfE\,  (\rho -\nabla\cdot  \bfE)\big)-\bfB\, (\nabla\cdot  \bfB)   \Big)d^3x\,, 
\label{bktLH}
\eq
in agreement with \eqref{conL}.

\medskip
\noindent\underline{$\{\bfM,H\}$}: 
\medskip

Using
\bqy
\frac{\de M_i}{\de f_s}&=&m_s\big(m_s \sqrt{1+ \abs{\bfv}^{2}} \, x_i -v_i\, t\big)\,,
\\
  \frac{\de M_i}{\de E_\ell}&=& E_\ell x_i -  \ep_{i\ell s} B_s t\,,
  \\
  \frac{\de M_i}{\de B_\ell}&=&B_\ell x_i - \ep_{is\ell} E_s t \,.
\eqy
we obtain
 \bq
\{ \bfM,H\}= t  \int \big(\bfB (\nabla\cdot  \bfB) - \bfE\,  (\rho -\nabla\cdot  \bfE)\big)\,  d^3x\,, 
\eq
in agreement with \eqref{COMM}.

\subsection{Relativistic Pairwise Brackets}
\label{ssec:RPBs}

Now consider the pairwise Poisson brackets of elements of $\{\bfP,\bfL, \bfM\}$.  As noted above, for each of these it is convenient to  break up the bracket into ``kinetic" and "field" terms.  For example,  with $\bfP=\bfP^K + \bfP^F$, the bracket
$ \{\bfP,\bfP\}$ becomes 
\bqy
\{P_i,P_j\}&=&\{P^K_i,P^K_j\} + \{P^K_i,P_j^F\}
\\
&&\hspace{2cm}+ \{P^F_i,P^K_j\}+ \{P^F_i,P^F_j\}
\nonumber
\eqy
and similarly for all the other pairs.  By direct calculation we obtain the following:
 
\medskip
\noindent\underline{$\{\bfP,\bfP\}$}: 
\medskip

\bqy
\{P^K_i,P^K_j\} &=&\int \rho \,\ep_{ijt} B_t\, d^3x
\label{KK}
\\
 \{P^K_i,P_j^F\} &+& \{P^F_i,P^K_j\} = -2\int \rho\, \ep_{ijt} B_t\, d^3x
 \label{KF}
\\
 \{P^F_i,P^F_j\}&=& \int \ep_{i\ell t} \ep_{\ell rk}\ep_{jsr} B_t \p_r E_s\, d^3x - (i\leftrightarrow j)
 \nonumber\\
 &=&\int\big( \ep_{ijt} B_t \nabla\cdot \bfE   - \ep_{ist} B_t  \p_j E_s
 \nonumber\\
 &&\hspace{.38cm} - \ep_{jit} B_t \nabla\cdot \bfE + \ep_{jst} B_t \p_i E_s\big)d^3x
 \nonumber
 \\ 
 &=&\int\big( \ep_{ijt} B_t \nabla\cdot \bfE  -  \ep_{ijt} E_t \nabla\cdot \bfB\big) d^3x\,,
 \label{lemma}
 \eqy
 where the last equatity of \eqref{lemma} follows from an identity proven in \cite{pjm13} (see Eq.~(A14)).  Summing  \eqref{KK}, and \eqref{KF}, and \eqref{lemma} yields
 \bq
\{P_i,P_j\}=\int\ep_{ijt} \big(  B_t( \nabla\cdot \bfE-\rho) -   E_t \nabla\cdot \bfB\big)\, d^3x\,.
\label{PP}
\eq

\medskip
\noindent\underline{$\{\bfL,\bfP\}$}:
\medskip

By direct calculation
\bqy
\{L_i^K,P_j^K\}&=& \ep_{ijk} P^K_k\!  + \! \int \! \rho\, ( \bfx\cdot\bfB\, \de_{ij} -  B_i x_j)\, d^3x \  \ 
\label{LKPK}\\
\{L^K_i,P_j^F\} &+& \{L^F_i,P^K_j\} =
\nonumber
\\
&-&  2  \int \! \rho\, ( \bfx\cdot\bfB\, \de_{ij} -  B_i x_j)\, d^3x 
\label{LPKF}\\ 
 \{L^F_i,P_j^F\} &=&\int (\bfx\cdot\bfB\, \de_{\ell j} -B_i x_\ell)\ep_{\ell s t}\p_s\ep_{jrt} E_r\, d^3x
\nonumber\\
 &&\hspace{ 2 cm} -\,  (\bfE\leftrightarrow\bfB)
  \label{LFPFa}\\
 &=&\ep_{ijk} L^F_k + \int (\bfx\cdot\bfB\, \de_{ij} -  B_i x_j  )\nabla\cdot \bfE \, d^3x
 \nonumber \\
 && - \int (\bfx\cdot\bfE\, \de_{ij} -  E_i x_j  )\nabla\cdot \bfB \, d^3x\,.
 \label{LFPF}
\eqy
Note, in obtaining \eqref{LFPF} the first and second terms of \eqref{LFPFa} are, respectively, 
\bqy
&&   -\int  B_i x_\ell        \ep_{\ell s t}\p_s\ep_{jrt} E_r\, d^3x =
\\
&&\hspace{1cm}  - \int ( B_i x_j \nabla\cdot \bfE + B_i E_j + \bfx\cdot\bfE\, \p_j B_i)\, d^3x
\nonumber\\
&&-\int \bfx\cdot\bfB\, \de_{\ell j}     \ep_{\ell s t}\p_s\ep_{jrt} E_r\, d^3x=
\\
&&\hspace{ 1cm}  - \int( \bfx\cdot\bfB\,  \de_{ij} \nabla\cdot\bfE  - \bfx\cdot\bfB\,   \p_jE_i )\, d^3x\,.
\nonumber
\eqy
Summing \eqref{LKPK}, \eqref{LPKF}, and \eqref{LFLF} yields
\bqy
\{L_i,P_j\}&=& \ep_{ijk} P_k   - \int (\bfx\cdot\bfE \, \de_{ij} -  E_i x_j  )\nabla\cdot \bfB \, d^3x 
 \nonumber \\
 &+&  \int (\bfx\cdot\bfB\, \de_{ij} -  B_i x_j  )(\nabla\cdot \bfE-\rho) \, d^3x\,. 
 \label{LP}
\eqy

\medskip
\noindent\underline{$\{\bfL,\bfL\}$}:
\medskip

\bqy
\{L_i^K,L_j^K\}&=&\sum_s m_s \int \big[ \ep_{iuk} x_u v_k, \ep_{jst} x_s v_t\big] f_s\, \dmu 
\nonumber\\
&&
\hspace{1cm} + \int \rho\, B_t\, \ep_{trs} \ep_{iur} \ep_{jvs}\,  x_ux_v \, d^3x
\nonumber\\
&=& \ep_{ijk} L^K_k + \int \rho\, \bfB\cdot\bfx \,\ep_{ijk} x_k \, d^3x
\label{LKLK}
\\
 \{L^K_i,L_j^F\} &+& \{L^F_i,L^K_j\} =-2 \int \rho\, \bfB\cdot\bfx \,\ep_{ijk} x_k \, d^3x
\label{LKLF}
\\
 \{L^F_i,L_j^F\} &=&\ep_{ijk} L^F_k 
  \label{LFLF}\\
&+& \int\ep_{ijk} x_k \big( \bfx\cdot\bfB\,  \nabla\cdot\bfE - \bfx\cdot\bfE\,  \nabla\cdot\bfB\big)d^3x\,.
\nonumber
\eqy
Summing \eqref{LKLK}, \eqref{LKLF}, and \eqref{LFLF} yields
\bqy
\{L_i,L_j\} &=&\ep_{ijk} L_k 
  \label{LL}\\
&+& \int\ep_{ijk} x_k \big( \bfx\cdot\bfB \,( \nabla\cdot\bfE-\rho) - \bfx\cdot\bfE\, \nabla\cdot\bfB\big)d^3x\,. 
\nonumber
\eqy

\medskip
\noindent\underline{$\{\bfM,\bfP\}$}:
\medskip

\bqy
\{M_i^K,P_j^K\}&=& H^F \de_{ij}  -   \int x_i (\bfJ\times\bfB)_j\, d^3x  
\label{MKPK}\\
\{M^K_i,P_j^F\} &+& \{M^F_i,P^K_j\} =
\nonumber
\\
&&    \int  \big(x_i (\bfJ \times \bfB)_j - \rho\,x_i E_j\big)\, d^3x 
\label{MKPF}\\ 
 \{M^F_i,P_j^F\} &=& H^F \de_{ij}
 \label{MFPF}\\
 &+&
 \int x_i(E_j \nabla\cdot\bfE - B_j \nabla\cdot\bfB)\, d^3x\,.
\nonumber
\eqy
Summing \eqref{MKPK}, \eqref{MKPF}, and \eqref{MFPF} yields
\bqy
\{M_i,P_j\}&=& H \de_{ij} 
\label{MP}\\
&+&   \int x_i\big(E_j (\nabla\cdot\bfE -\rho) - B_j \nabla\cdot\bfB\big)\, d^3x \,.
\nonumber
\eqy

\medskip
\noindent\underline{$\{\bfM,\bfM\}$}:
\medskip

Using $\bfM= \bfX-\bfP t$ and 
\[
\{M_i,M_j\} = -\{P_i t,M_j\} + \{M_i, P_j t\} + \{X_i,X_j\}\, ,
\]
direct calculation yields
\bqy
\{X_i^K,X_j^K\}&=& -\ep_{ijk}L_k^K
\label{MKMK}\\
\{M_j, P_i t\} &+& \{M_i, P_j t\}  = t \!\!\int\big(( x_i E_j - x_j E_i)(\nabla\cdot\bfE -\rho) 
\nonumber\\
&-&  ( x_i B_j - x_j B_j )(\nabla\cdot\bfB)\big)\, d^3x 
\label{PMt}\\ 
 \{X^F_i,X_j^F\} &=& -\ep_{ijk}L^F_k\,.
 \label{XFXF}
\eqy
Summing \eqref{MKMK}, \eqref{PMt}, and \eqref{XFXF} yields
\bqy
\{M_i,M_j\}&=& - \ep_{ijk}L_k + t \int\big(( x_i E_j - x_j E_i)(\nabla\cdot\bfE -\rho) 
\nonumber\\
&-&  ( x_i B_j - x_j B_j )(\nabla\cdot\bfB)\big)\, d^3x \,.
\label{MM}
\eqy

\medskip
\noindent\underline{$\{\bfL,\bfM\}$}:
\medskip

\bqy
\{L_i,M_j\}&=& \{L_i,X_j\} - \{L_i, P_j\}t
\nonumber\\
&=&  \{L_i,X_j\} -  \ep_{ijk}P_k t
\label{LMKP}\\
\{L_i^K,X_j^K\}&=& \ep_{ijk} X_k^K
\label{LKXK}
\\  
&+&   \int \big(J_i x_j \,\bfx\cdot\bfB - x_j B_i\,\bfx\cdot\bfJ\big)\, d^3x  
\nonumber\\
\{L^K_i,X_j^F\} &+& \{L^F_i,X^K_j\} = \int \big(\rho\, \ep_{ir\ell} x_r x_j E_{\ell}\big)\, d^3x  
\nonumber
\\
&-& \int \big(J_i x_j \,\bfx\cdot\bfB - x_j B_i\,\bfx\cdot\bfJ\big)\, d^3x  
\label{MKXF}\\ 
 \{L^F_i,X_j^F\} &=&   \ep_{ijk} X^F_k  +  \int \big(\ep_{rik} B_k x_j x_r  \nabla\cdot\bfB  
\nonumber\\
 &+&
\ep_{rik} E_k x_j x_r \nabla\cdot\bfE\big)\, d^3x \,.
 \label{LFXF}
\eqy

Using \eqref{LMKP} and summing \eqref{LKXK}, \eqref{MKXF}, and \eqref{LFXF} yields
\bqy
\{L_i,M_j\}&=&  \ep_{ijk} M_k  +  \int \big(\ep_{rik} B_k x_j x_r  \nabla\cdot\bfB 
\label{LM}\\
&+&  \ep_{rik} E_k x_j x_r (\nabla\cdot\bfE -\rho) \big)\, d^3x \,.
\nonumber
\eqy

\subsection{Nonrelativistic  Brackets}
\label{ssec:I}

From Sec.~\ref{ssec:RI} it is evident that  nonrelativistic invariants $\hat\bfP$ and $\hat\bfL$ are identical to those of \eqref{Pinv} and \eqref{Linv}, respectively, and that $\{\hat\bfP,\hat H\}$ and $\{\hat\bfL, \hat H\}$ are given by expressions identical  to  \eqref{bktPH} and \eqref{bktLH}.  However,  there is no invariant $\hat\bfM$,  analogous to the $\bfM$ of \eqref{bfM},  for which  $\{\hat M,\hat H\}=0$.  As noted in Sec.~\ref{ssec:Ccom} this is because  the Vlasov-Maxwell system has neither full Galilean nor full Poincar\'{e} symmetry.     

Because the nonrelativistic invariants are identical to the relativistic and because both theories are generated by the same bracket \eqref{VMbracket}, the  nonrelativistic pairwise brackets are clearly identical to those of  \eqref{PP}, \eqref{LP},  and \eqref{LL}.

\subsection{Lie algebra realization of invariants}
 
Finite-dimensional Lie algebras in terms of matrices are Lie algebra representations, while infinite-dimensional Poisson brackets defined on either functions or functionals are realizations (see e.g.\ \cite{sudarshan}), e.g., the bracket of \eqref{VMbracket} defined on the set of all functionals of $\Psi=\{f_s,\bfE,\bfB\}$ is  closed, bilinear, antisymmetric, and satisfies the Jacobi identity (provided $\bfB$ is solenoidal). 
 
As noted at the beginning of the present section, a  subalgebra of a Poisson bracket realization composed of the functionals that satisfy $\{H,\Phi\}=0$  is an algebra of invariants, and this subalgebra is closed because of Poisson's theorem. 

 The results of Secs.~\ref{ssec:RI} and \ref{ssec:I} show that for the Vlasov-Maxwell  realization with bracket \eqref{VMbracket} the algebra of invariants has a complicated structure because of terms containing $\nabla\cdot \bfB$ and
  $\nabla\cdot \bfE-\rho$.   Because of the Jacobi identity the $\nabla\cdot \bfB$ should be posthaste set to zero to insure that we have an algebra.  Moreover, from Sec.~\ref{sec:DandF} we see all quantities $\bfP, \bfL$, and $\bfM$ require in addition   $\nabla\cdot \bfE-\rho =0$ in order to in fact be invariants.  Thus upon restricting to the Casimir leaf, i.e., constraint subspace with  $\nabla\cdot \bfB=0$ and $\nabla\cdot \bfE-\rho =0$, we obtain the following Lie algebra of invariants, for the relativistic case
\bqy
\{P_i,P_j\}&=& 0\\
\label{PPla}
\{L_i,P_j\}&=& \ep_{ijk} P_k\\
\{L_i,L_j\}&=& \ep_{ijk} L_k\\
\{M_i,P_j\}&=& \de_{ij} H\\
\{M_i,M_j\}&=&- \ep_{ijk} L_k\\
\{L_i,M_j\}&=& \ep_{ijk} M_k
\label{LMla}
\label{Rnoncom}
\eqy
with $\{H,\Phi\}=0$.  A physicists then immediately recognizes Eqs.~\eqref{PPla}--\eqref{LMla} to be a realization of the  ten parameter Poincar\'{e} Lie algebra possessed by  relativistic  field theories.  (Note, sometimes $\{X_i,H\}= P_i$ is used instead of $\{M_i,H\}= 0$.)

The elements of the Poincar\'{e} algebra are  generators of the various infinitesimal transformations.  For example, $\{\Psi,H\}$ generates infinitesimal time translations of our field dynamical variables, as is evident from the Hamiltonian formulation of Sec.~\ref{sec:vlasov-maxwell}, while  infinitesimal space translation of the distribution functions are generated  via 
\bq
\{f_s ,P_i\} = -\p_i f_s\,.
\eq
In this way the the entire Poincar\'e group, composed of a semidirect product of  the subgroup of homogeneous Lorentz transformations  generated by $\bfL$ and  $\bfM$ with space-time translations generated by $\bfP$ and $H$, generates transformations of our field variables.  See,  e.g., \cite{ibb} where this is worked out for the source-free Maxwell  equations and  \cite{ibb2} where they were first worked out for the Vlasov-Maxwell system.  Also note, a representation with generators for the Gallilean group  in the context of MHD can be found in  \cite{pjm82}. 
 
The above Lie algebra expressions, in fact the entire theory \cite{pjmMMT86} can be generalized to covariant form with a 4-metric $g_{\mu\nu}$ and to arbitrary 3-space coordinates by writing in the Hamiltonian $|\bfv|^2= v^iv_i$, $|\bfE|^2= E^iE_i$, and $|\bfB|^2= B^iB_i$ where a 3-metric $g_{ij}$ appropriate to the coordinates is used to raise and lower indices.  All can also be cast into geometric form, but this is beyond the present scope.

For the nonrelativistic case,  restriction  to the Casimir leaf with  $\nabla\cdot \bfB=0$ and $\nabla\cdot \bfE-\rho =0$ produces the  following Lie algebra of invariants
\bqy
\{\hat{P}_i,\hat{P}_j\}&=& 0\\
\{\hat{L}_i,\hat{P}_j\}&=& \ep_{ijk} \hat{P}_k\\
\{\hat{L}_i,\hat{L}_j\}&=& \ep_{ijk} \hat{L}_k\,,
\eqy
together with $\{H, \bfL\}=\{H,\bfP\}=0$.   Thus we have a realization of the algebra associated with the Euclidean group of translations and rotations, but as noted before the full ten parameter Galilean group does not occur.  In a manner similar to the relativistic case, these invariants in the Poisson bracket of \eqref{VMbracket} produce infinitesimal transformations of the  field variables $\Psi$.

\section{Discussion}
\label{sec:discussion}

\subsection{Frozen background ions}
\label{ssec:frozen}

 Consider  a single species plasma, say electrons,  with a neutralizing positive background charge $\rho_i(\bfx)$, say ions.
Observe $\rho_i(\bfx)$ may, but need not, be constant in space and  by  neutralizing we mean $\int (\rho_e-\rho_i)d^3x=0$, 
where $\rho_e(\bfx,t)$ is the time dependent electron density. 
For this case, Gauss's law  is given by 
 \bq
 \nabla\cdot \bfE= \rho_i -\rho_e\,, 
 \eq
and the Poisson bracket \eqref{VMbracket}   has the local Casimir invariant, equivalent to \eqref{CE}, of 
\bq
\calc_E=  \nabla\cdot \bfE +\rho_e\,,
\eq
as can be shown by direct substitution.  Observe that the Casimir does not recognize the existence of the background charge.  

Because the background is stationary, the total momentum is given by 
\bq
\bfP= \int m_e  \bfv\,  f_e \,   \dmu + \int   \bfE\times\bfB\, d^3x\,,
\eq 
and  according to \eqref{dP0} it satisfies 
 \bq
 \frac{d\bfP_e}{dt} = \{ \bfP_e, H\}= -\int \bfE(\bfx,t) \, \rho_i(\bfx)\, d^3x\,.
 \label{dotPe}
 \eq
Thus, even when evaluated on the constraints of \eqref{constraints} with flux conserving boundary conditions, this well-known system of plasma physics does not conserve momentum, even in the case where $\rho_i$ is constant in space as well as time.

The quantity $\bfE \, \rho_i(\bfx)$ is the force per unit volume on the ions caused by any electric field in the system. Since the ions are forced to be stationary,  the total force on the electrons as a  back-reaction is $-\int \bfE \, \rho_i(\bfx)\, d\bfx$.  For the usual case of spatially constant $\rho_i$, \eqref{dotPe} is
 \bq
 \frac{d\bfP_e}{dt} =  - \rho_i \int\! \bfE  \, d^3x\,.
 \label{dotPec}
 \eq
In \eqref{dotPec} various boundary conditions are possible and $\bfE$ may have longitudinal (electrostatic)  as well as transverse contributions. 

A similar state of affairs applies on $\nabla\cdot E-\rho=0$ for the conservation of angular momentum  and the other invariants.

\subsection{Computational consequences}
\label{ssec:consequences}

The connection between momentum and charge conservation is well known in the computational science of the Vlasov equation.  However, we amplify on this theme in this subsection, throughout which it will be assumed that $\nabla\cdot \bfB=0$ is exactly maintained. 

 Consider the case of Sec.~\ref{ssec:frozen} with the constant ion background density.   Assuming a spatial  $2\pi$-cube with  periodic boundary conditions,  and writing
 \bq
\bfE=\frac1{(2\pi)^3}\sum_{\bfk \in \Z^3} \bfE_{\bfk}\,  e^{i\bfk\cdot \bfx}\,,
\label{fourier}
\eq
which is often the case for Vlasov computations,   \eqref{dotPec} becomes 
\bq
\frac{d\bfP_e}{dt} =  -   \rho_i\,  \bfE_0\,,
\label{dotP0}
\eq
where $\bfE_0$ is the $\bfk=0$ Fourier component.  Thus,  if  $\bfE_0$ is removed from a code, it will conserve momentum -- if not there can be secular growth.

More generally, suppose in a computation Gauss's law is not maintained,  either because of an improper initial condition or because of numerical error, i.e.,  a  charge imbalance  $\De_{\rho}(\bfx)\coloneqq \rho -\nabla\cdot \bfE$ ensues.  If this happens,  \eqref{dP0} implies 
 \bq
 \frac{d\bfP}{dt} = \{ \bfP, H\}= \int \bfE \, \De_{\rho}(\bfx)\, d^3x 
 \label{dotP}
 \eq
and we have violation of momentum conservation.  For this reason codes such as GEMPIC \cite{pjmKKS17} that exactly conserve Gauss's law  (as well as $\nabla\cdot \bfB=0$) are desirable.  In  the Fourier PIC context of  \cite{evstatiev-shadwick, shadwick-stamm-evstatiev} this has been noted. Alternatively, enforcing momentum conversation in the GEMPIC context, i.e., using  finite element exterior calculus, can be achieved by  incorporating  Dirac constraint theory (see e.g.,  \cite{pjmLB09,pjmCGBT13}).   
 
Examination of the results of  Sec.~\ref{sec:DandF} shows that the feature above  not only occurs for momentum but for  other invariants as well.    Equation \eqref{conL} shows that conservation of the angular momentum $\bfL$ similarly relies on maintaining the constraints.  The Dirac constraint theory is a promising tool  for preserving the entire algebra of invariants in the GEMPIC framework.  Indeed  momentum conservation for Vlasov-Poisson system has been achieved by this method  \cite{esonnen24}, but it remains to be done for the full algebra of invariants.  Of particular interest might be  the invariant $\bfM$, which to our knowledge has not been monitored in any computation.   From \eqref{COMM} we see  the right-hand side has  the constraints with  a factor of  $t$, thus if these are not satisfied $\bfM$ may experience rapid growth.  Therefore, this invariant could be a particularly good diagnostic for assessing the accuracy of numerical algorithms.  However, again, for schemes like GEMPIC \cite{pjmKKS17} it can be exactly conserved using the Dirac method.

\subsection{Weibel system}
\label{ssec:weibel}

One of the simplest Vlasov-Maxwell systems is that used for study of the Weibel instability (e.g.\ \cite{pegoraro96,califano1998ksw,pjmCGL14,pjmKKS17}).  This system, which  assumes the dynamics takes place on a domain with one spatial variable $x_1$ and two velocity variables $v_{1,2}$, has the reduced  equations of motion 
\bqy
    \frac{\partial f}{\partial t} + v_1\frac{\partial f}{\partial x_1}  &-&  (E_1 + v_2 B_3) \frac{\partial f}{\partial v_1}  
 \label{weibelV}\\
 &&- \  (E_2 - v_1 B_3) \frac{\partial f}{\partial v_2}  =0\,,
 \nonumber
\eqy
and 
 \bqy
&& \frac{\partial E_1}{\partial t}= - J_1\,,  
\label{dotE1} \\
&&   \frac{\partial E_2}{\partial t}= -\frac{\partial B_3}{\partial x_1}- J_2\,, 
\\
 && \frac{\partial B_3}{\partial t}= -\frac{\partial E_2}{\partial x_1}\,.
 \label{Wmax}
\eqy
Here  
\bq
J_{1,2} =- \int v_{1,2}\,  f \, dv_1 dv_2\,, 
\eq
where dimensionless variables have been assumed.  
Observe, \eqref{dotE1} replaces Gauss's law, $\nabla\cdot \bfE = 1- \rho$, which  is not directly integrated but serves as an initial condition in this dynamics.

Assuming periodic boundary conditions,  expanding as in \eqref{fourier},  and  using $\langle \bfE \rangle = \bfE_0$,  \eqref{dotP0} becomes 
\bq
\frac{d{P_e}_1}{dt} =  -  \langle E_1\rangle  \qquad \mathrm{and}\qquad 
\frac{d{P_e}_2}{dt} =  -\langle E_{2}\rangle\,. 
\eq
and we see again, as in Sec.~\ref{ssec:frozen}, that momentum conservation can be violated because of the force of constraint to maintain the stationarity of the ion charge.  Even if  $\langle E_1\rangle =0$ at $t=0$, it need not stay there.  From \eqref{dotE1} we obtain
\bq
 \frac{\partial  \langle E_1\rangle }{\partial t}= - \langle J_1 \rangle \,,
 \eq
and in  general $\langle J_1 \rangle\not \equiv 0$ so momentum is not conserved.   However, as  a means for  testing momentum conservation one can use the approach of \cite{Crouseilles:2015} and replaced \eqref{dotE1} by
\bq
 \frac{\partial E_1}{\partial t}= - J_1 +  \langle J_1 \rangle \,.  
 \eq
 which is not a physical equation.   Nevertheless,  in  this artificial dynamics momentum will be conserved if $\langle E_1\rangle =0$ at $t=0$.  A similar artifice may be possible in other situations. 
 
\subsection{Constraint asymmetry}
\label{ssec:assymetry}

As has been noted several times in this paper, the constraints $\nabla\cdot\bfB=0$ and $\nabla\cdot\bfE-\rho=0$  do not play the same roles in the Hamiltonian framework of the Vlasov-Maxwell system.  While Gauss's law  is a bona fide  Casimir invariant,  $\nabla\cdot\bfB=0$ is  necessary for satisfying the Jacobi identity and therefore is essential for  the existence of the  Hamiltonian structure.  This   property  was unknown prior to \cite{pjm82} and appears to not have been fully appreciated.  It is  a general property of the full Maxwell equations coupled to dynamical matter fields, and a version of it even exists in  Dirac's theory of monopoles \cite{pjm13,pjmH18}.

In the context of magnetohydrodynamics,  a noncanonical Poisson bracket  independent of   $\nabla\cdot\bfB=0$ was obtained in \cite{pjmG82}, but  despite considerable effort it does not appear to be possible to remove it for the Vlasov-Maxwell system.  However, using  Dirac constraint theory it was shown in \cite{pjmCGBT13}  that  $\nabla\cdot\bfB=0$ can be reduced  to a boundary condition at infinity, but this amounts to usual proof that demonstrates closed and  exact differential forms are equivalent  on $\R^3$. Considerable effort has been spent on determining the validity of the Bohm-Aharnov effect, both theoretically and experimentally, where the $\nabla\cdot\bfB=0$ issue is at the heart of the matter.  In \cite{batelaan09} it was pointed out that experimental verification of the electric Bohm-Aharonv effect has not achieved the same level of credibility as the magnetic.  Perhaps this distinction between $\nabla\cdot\bfB=0$ and $\nabla\cdot\bfE-\rho=0$, which we have pointed out,  is the essence of this difference.   Further pursing this is beyond the scope of the present paper. 

\section{Summary}
\label{sec:summary}
 
 In this paper we have reviewed the Hamiltonian structure of the Vlasov-Maxwell system, both nonrelativisitic and relativistic.  The Hamiltonian for each case and the common  Poisson bracket and associated Casimir invariants were given.  Conservation laws for both cases were given and their associated fluxes.  The algebra of invariants, determined by the invariants and the Poisson bracket,  was obtained and shown to be    realization of the algebra associated with the Poincar\'e group in the relativistic case and the Euclidean group in the nonrelativistic case.  In demonstrating the above particular care was given to the roles played by  the constraints of Gauss's law and the monopole condition, where the asymmetry in these constraints, as first pointed out in \cite{pjm82},  was described.  The special cases of the frozen ion background and the Weibel system were described.   The development was given with an eye toward possible computational ramifications.

\section*{Acknowledgment}
\noindent    
The author received support from the  U.S. Dept.\ of Energy Contract \# DE-FG02- 04ER54742 and from a Forschungspreis from  the Alexander von Humboldt Foundation.  He would like to  warmly  acknowledge the hospitality of the Numerical Plasma Physics Division of the IPP, Max Planck, Garching.





\bibliographystyle{apsrev}



\end{document}